\pdfoutput=1
\documentclass[conference]{IEEEtran}
\IEEEoverridecommandlockouts
\usepackage{cite}
\usepackage{amsmath,amssymb,amsfonts}
\usepackage{algorithmic}
\usepackage{graphicx}
\usepackage{textcomp}
\usepackage{xcolor}
\def\BibTeX{{\rm B\kern-.05em{\sc i\kern-.025em b}\kern-.08em
    T\kern-.1667em\lower.7ex\hbox{E}\kern-.125emX}}

\usepackage[hyphens]{url}
\usepackage{subcaption}
\begin{document}

\newcommand\copyrighttext{
	\Huge {IEEE Copyright Notice} \\ \\
	\large {Copyright (c) 2019 IEEE \\
		Personal use of this material is permitted. Permission from IEEE must be obtained for all other uses, in any current or future media, including reprinting/republishing this material for advertising or promotional purposes, creating new collective works, for resale or redistribution to servers or lists, or reuse of any copyrighted component of this work in other works.} \\ \\
	
	{\Large Accepted to be published in: 2019 IEEE 16th International Conference on Ubiquitous Intelligence and
		Computing (IEEE UIC 2019), August 19-23, 2019} \\ \\ 
	
	\vspace{2cm}
	
	Cite as:\\
	\includegraphics[trim={2.075cm 23cm 9cm 3.1cm},clip]{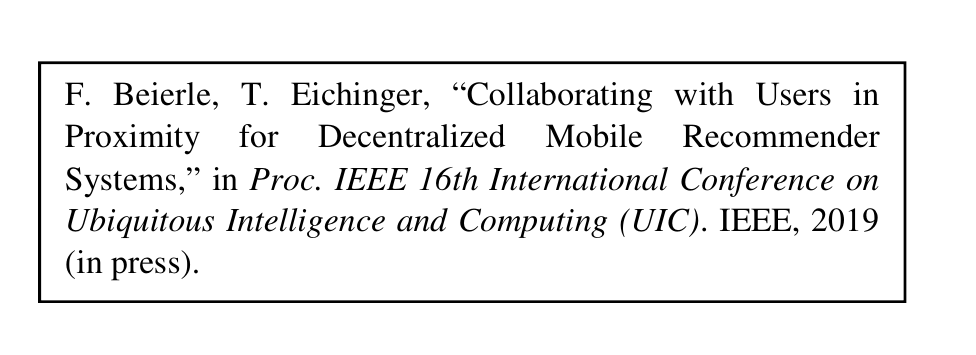}
	
	\vspace{1.5cm}
	
	BibTeX:\\ \\
	\includegraphics[trim={2.25cm 21cm 2cm 4cm},clip]{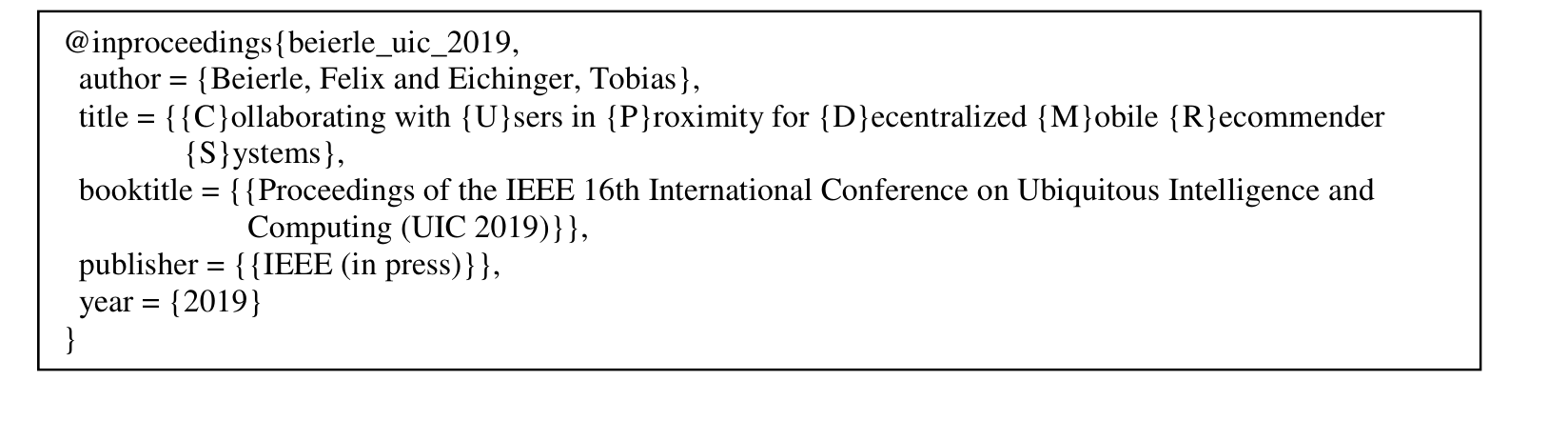}
}

\twocolumn[
\begin{@twocolumnfalse}
	\copyrighttext
\end{@twocolumnfalse}
]

\title{Collaborating with Users in Proximity for Decentralized Mobile Recommender Systems}

\author{\IEEEauthorblockN{Felix Beierle}
	\IEEEauthorblockA{\textit{Service-centric Networking}\\
		\textit{Telekom Innovation Laboratories}\\
		\textit{Technische Universität Berlin}\\
		Berlin, Germany\\
		beierle@tu-berlin.de}
	\and
	\IEEEauthorblockN{Tobias Eichinger}
	\IEEEauthorblockA{\textit{Service-centric Networking}\\
		\textit{Telekom Innovation Laboratories}\\
		\textit{Technische Universität Berlin}\\
		Berlin, Germany\\
		tobias.eichinger@tu-berlin.de}
}

\maketitle

\begin{abstract}
Typically, recommender systems from any domain, be it movies,
music, restaurants, etc., are organized in a centralized fashion.
The service provider holds all the data,
biases in the recommender algorithms are not transparent to the user,
and the service providers often create lock-in effects making it
inconvenient for the user to switch providers.
In this paper, we argue that the user's smartphone already holds
a lot of the data that feeds into typical recommender systems
for movies, music, or POIs.
With the ubiquity of the smartphone and other users in proximity
in public places or public transportation,
data can be exchanged directly between users in a device-to-device manner.
This way, each smartphone can build its own database
and calculate its own recommendations.
One of the benefits of such a system is that it is not restricted
to recommendations for just one user -- ad-hoc group recommendations
are also possible.
While the infrastructure for such a platform already exists -- the
smartphones already in the palms of the users -- there
are challenges both with respect to the mobile recommender system
platform as well as to its recommender algorithms.
In this paper, we present a mobile architecture for the described
system -- consisting of data collection, data exchange, and recommender system --
and highlight its challenges and opportunities.

\end{abstract}


\section{introduction}
\label{sec:intro}

Recommender systems are ubiquitous in several different domains of everyday life.
They recommend restaurants to go to (e.g., Yelp, Google Maps),
music to listen to (e.g., Spotify, Deezer),
or movies to watch (e.g., IMDb, Netflix).
The recommender algorithms usually are limited to recommending items
that the platform offers (Spotify, Netflix) or at least items that are
indexed with the provider (Yelp, Google Maps).
There might be certain biases, for example, towards recommending products that
create the biggest margin for the service provider.
Furthermore, as the service provider is interested in retaining its users,
there are certain lock-in effects that try to make the user stay with the
current service provider instead of switching to another one.
In that sense, from an organizational perspective, the examples given above
are centralized -- a single service provider has all the data and decides
how recommendations are calculated.

On top of that, users have privacy concerns
regarding the use of the data they share \cite{FalchBusinessmodelssocial2009}.
Recently, Google was fined 50 million euros due to a violation of the new European privacy laws (GDPR)\footnote{\url{https://www.cnil.fr/en/cnils-restricted-committee-imposes-financial-penalty-50-million-euros-against-google-llc}}.

The infrastructure that could be a solution for both challenges
of lock-in effects and privacy concerns at the same time, is already
in the palms of its users.
The smartphone can store lots of information about its user and his/her interests, e.g., regarding preferred restaurants, music, or movies.
Equipped with capabilities for device-to-device communication,
users can exchange data between each other.
When considering recommender systems based on
content-based filtering or collaborative filtering,
data about similar items and similar users is needed.
Data about the properties of items can be retrieved through public APIs
(e.g., Google Places, Spotify, Open Movie Database (OMDb)).
Finding similar users might be simple with smartphones:
spending time at the same location might imply similarity -- at least to a certain degree.
Additionally, from our previous research, other methods of
determining similarity between users based on smartphone data are
available \cite{BeierleYouWhatSimilarity2018,EichingeraffinitySystemLatent2019}.
Thus, exchanging data between smartphones in proximity in a
device-to-device fashion allows to create local databases that allow to filter
for similar users.
This data can be used for on-device recommender systems that are
independent of external service providers.

Typically, each of the mentioned existing recommender systems
only offers recommendations for a single user.
The ubiquitous system we propose in this paper
offers the possibility of spontaneous ad-hoc
recommendations for groups of users in proximity.

Combining and expanding approaches from device-to-device computing
\cite{AhmedSurveySociallyAware2018,ZHANGcollaborativemultidevicecomputing2018}
and decentralized recommender systems
\cite{ZieglerSemanticWebRecommender2005a,Baragliapeertopeerrecommendersystem2013,BarbosaDistributedUserBasedCollaborative2018},
in this paper, we propose a modular architecture
for recommender systems
for virtually any domain,
building on the existing infrastructure of smartphones.
The architecture consists of collaborative data collection
paired with data exchange via device-to-device communication
and local recommender systems running on each device,
supported by third-party service providers where appropriate.
There are some challenges to overcome when developing such a platform.
Some data is already readily available on smartphones,
for example the most frequently visited locations.
Other user preferences/ratings that cannot be assessed automatically
might have to be entered manually or retrieved from external service
providers, e.g., music listened to or favorite movies.
While some short-distance wireless technologies, like NFC, Bluetooth,
or WiFi Direct, are available on most modern smartphones,
data exchange between users remains a challenge
to be implemented for multiplatform apps.

The main contributions of this paper are:
\begin{itemize}
	\item proposing a general modular architecture for a service-provider-independent mobile platform for recommender systems
	\item developing a comprehensive approach for collecting data and exchanging data
	in a device-to-device fashion for multiplatform apps (Android and iOS) to
	enable domain-independent recommender systems
\end{itemize}

The remainder of this paper is structured as follows:
In Section \ref{sec:architecture}, we introduce a general architecture
for the proposed mobile recommender system platform.
In Section \ref{sec:collecting-data} and \ref{sec:exchanging-data},
we illustrate our approach for collecting and exchanging data.
In Section \ref{sec:recommending}, we highlight
what challenges and opportunities the recommender systems in our proposed platform
will face, before pointing out related work in Section \ref{sec:relatedwork}.

\section{Proposed Architecture}
\label{sec:architecture}

In order for the described architecture to be feasible,
there are two main requirements:
\begin{enumerate}
	\item[R1] The system has to be multiplatform.
	\item[R2] The system has to be designed in a modular way.
\end{enumerate}
Android and iOS are the remaining relevant mobile operating systems.
They have a combined market share of
about 100\%\footnote{\url{https://www.statista.com/statistics/266136/global-market-share-held-by-smartphone-operating-systems/}}.
The main challenge this brings is related to data exchange,
as we describe in Section \ref{sec:exchanging-data}.
The system should be developed in a modular way (R2)
in order to be able to exchange components easily.
Consider short-distance wireless interfaces:
In the past, infrared was used.
Technological advances now offer larger transmission ranges, shorter connection times, and higher bandwidths via, for example, Bluetooth or WiFi Direct.
Similarly, advances in recommender systems
and machine learning might offer better
recommendations, creating the need to replace the module
or offload certain tasks to components available from
external service providers.

\begin{figure}[h]
	\includegraphics[width=1\columnwidth]{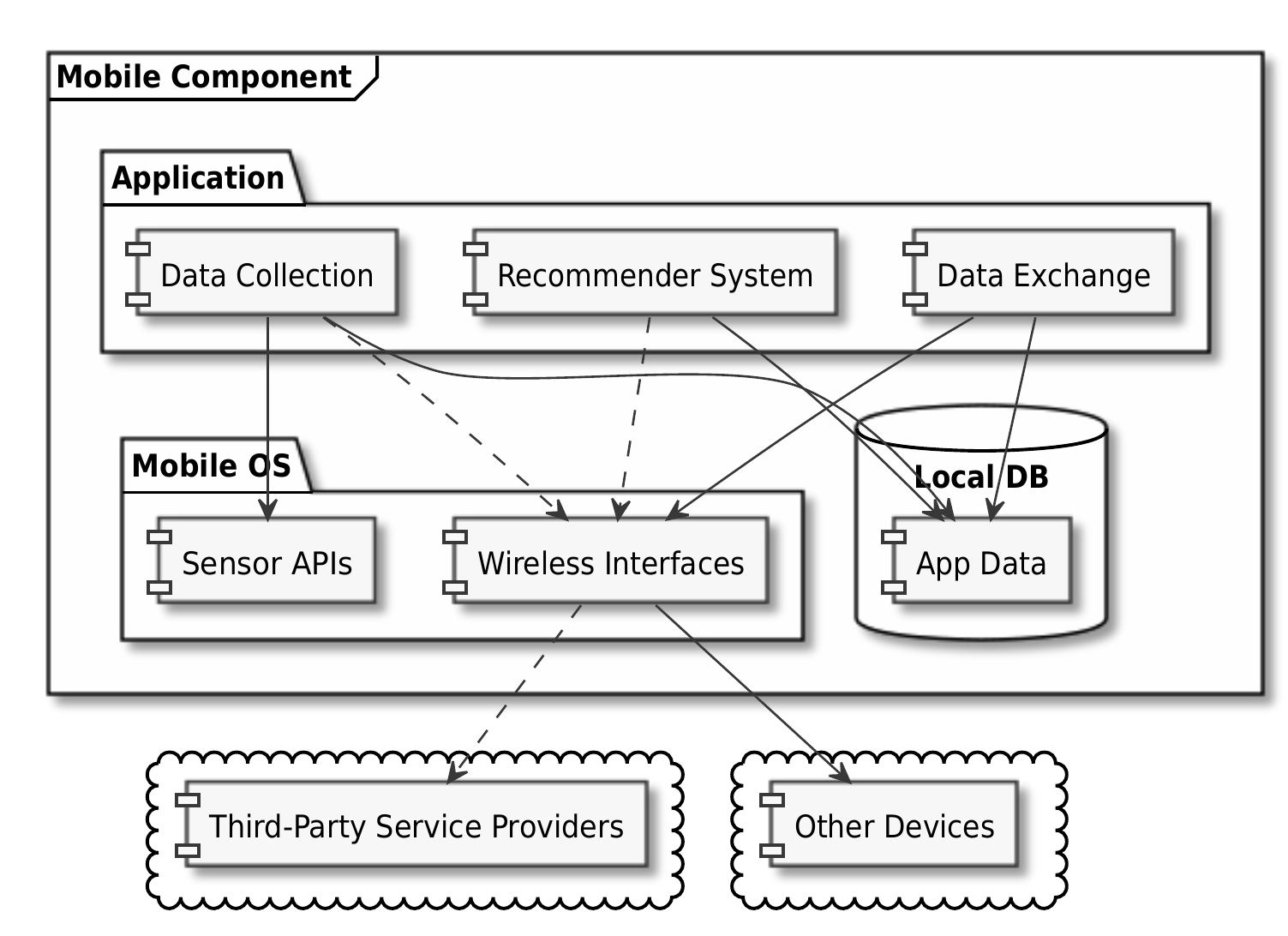}
	\caption{Architecture components of the proposed system.}
	\label{fig:architecture}
\end{figure}

In Figure \ref{fig:architecture},
we illustrate our proposed general modular architecture.
The three main components of the system are
\emph{Data Collection}, \emph{Data Exchange}, and \emph{Recommender System}.
Data Collection is responsible for getting data about the user.
Data Exchange is responsible for getting data from other users.
The Recommender Systems utilizes all available data
for recommending items to the user.
The mobile OS provides components for
sensors (for example for tracking the user's location for inferring
his/her favorite POIs) and wireless interfaces (for exchanging data).

As we further analyze in Section \ref{sec:collecting-data} and \ref{sec:exchanging-data},
external service providers might be needed (or be useful) in order to
retrieve metadata about items,
utilize existing systems, or offload data or computational tasks.
Figure \ref{fig:architecture} shows dashed lines for optional
connections to third party service providers.
Data Collection might use this to retrieve data about the user
or to enrich already available data, e.g., find out the genre
of the songs the user listened to.
More details are given in Section \ref{sec:collecting-data}.
The Recommender System can optionally be relayed to
an external service provider.

\section{Collecting Data}
\label{sec:collecting-data}

We identify three different possibilities to retrieve user data:

\begin{figure*}[ht]
	\centering
	\begin{subfigure}{0.355\columnwidth}
		\rule[-5mm]{0mm}{0mm}\includegraphics[width=\linewidth]{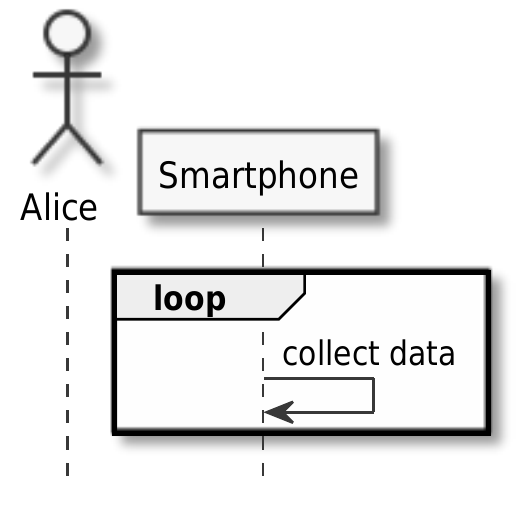}
		\caption{Automatic context data tracking.}
		\label{fig:data-collection-tracking}
	\end{subfigure}
	\qquad
	\begin{subfigure}{0.665\columnwidth}
		\rule[0mm]{0mm}{0mm}\includegraphics[width=\linewidth]{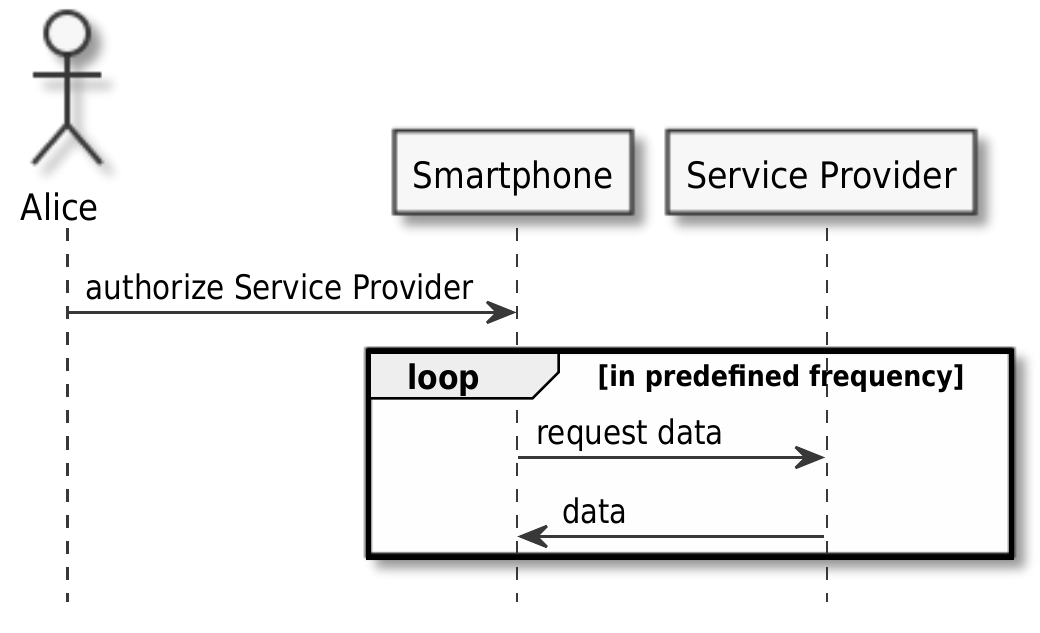}
		\caption{Retrieving data from third-party service providers.}
		\label{fig:data-collection-third-party}
\end{subfigure}
	\qquad
	\begin{subfigure}{0.665\columnwidth}
		\rule[0mm]{0mm}{0mm}\includegraphics[width=\linewidth]{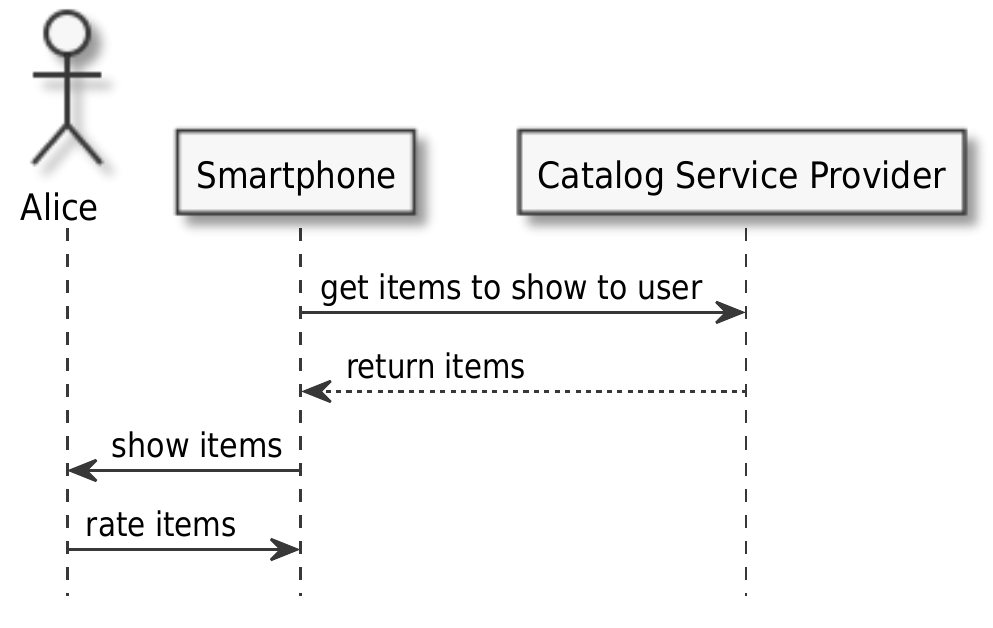}
		\caption{Letting the user manually rate items.}
		\label{fig:data-collection-manual}
	\end{subfigure}
	\caption{Data collection mechanisms.}
	\label{fig:data-collection}
\end{figure*}

\paragraph{Tracking data automatically}

Smartphones contain a multitude of sensors
that are often used for context-aware applications.
Frameworks like the Google Awareness API\footnote{\url{https://developers.google.com/awareness/}}
yield the location, weather, etc.\ for the user.
Such data can easily be tracked
and so preferences or implicit ratings for, e.g., locations or POIs can be inferred.
In previous work, we demonstrated an Android application
that tracks a large variety of context data sources
\cite{BeierleTYDRTrackYour2018,BeierleContextDataCategories2018,BeierleJAIHC2019}.
The data that can be tracked automatically on iOS might differ.
In order to create a multiplatform system and ensure
that the same data points are available on all systems,
additional ways of retrieving the user's ratings are necessary.
Figure \ref{fig:data-collection-tracking} shows the sequence diagram
of automatic context data tracking.

\paragraph{Retrieving data from existing service providers}

In order to minimize necessary user effort,
the second method we suggest is retrieving data from existing service providers.
For example, Spotify's API enables application developers to 
fetch recently played
tracks\footnote{\url{https://developer.spotify.com/documentation/web-api/reference/player/get-recently-played/}}.
Regularly doing this yields a complete music listening history
indicating implicit user ratings.
Figure \ref{fig:data-collection-third-party} shows the sequence diagram
for the collection of data from a third-party service provider.

\paragraph{Letting the user manually rate items}

For data that is neither automatically trackable nor available
via third parties, the user should be able to enter it manually.
By defining an ontology for categories and terms that can be exchanged between users,
compatibility between the data from different collection methods can be ensured.
Pre-defined categories can be movies, music, or restaurants, where recommender
system are often used, but any other category would be possible too.
Service providers like
The Open Movie Database API\footnote{\url{http://www.omdbapi.com/}},
for example, can be used to help
employ
globally valid identifiers for each item, in this case, for each movie.
Figure \ref{fig:data-collection-manual} shows the sequence diagram
for manual data collection.
\emph{Catalog Service Provider} denotes a service provider that offers structured
information about a specific category, like the mentioned Open
Movie Database.

\section{Exchanging data}
\label{sec:exchanging-data}

The idea behind the exchange of data is that when users pass each other,
their preference data, i.e., their ratings, are exchanged automatically in a device-to-device manner,
building up each user's local database with more data.
In order for such a system to work unobtrusively
and without user interaction,
the exchange of data should be done in the background (\emph{broadcasting}),
without establishing explicit connections between smartphones.
While device-to-device communication has gained some attention in research,
practical application is still lacking, especially when
considering broadcasting, and especially when
considering R1 (multiplatform app for both Android and iOS).
In this section, we give an overview of related work,
related applications, and propose a technical solution
to be used in our proposed architecture.

During the advent of mobile phones, researchers suggested using
device-to-device peer discovery and communication
in order to stimulate social interactions.
There are several papers between 2005 and 2010 describing
exchanging data via Bluetooth, sometimes combining
the direct data exchange with retrieving data from a central server
\cite{BeachWhozThatEvolvingEcosystem2008,EagleSocialSerendipityMobilizing2005,PietilainenMobiCliqueMiddlewareMobile2009,YangESmallTalkerDistributedMobile2010}.
In both \cite{ManweilerSMILEencounterbasedtrust2009} and \cite{TengEShadowLubricatingSocial2014},
the authors suggest using WiFi SSIDs and the Bluetooth discovery protocol
in order to exchange data between devices.
Only very small amounts of data can be transferred that way
but the benefit is that no proper connection has to be established between
two devices, thus allowing for broadcasting data to devices in proximity.

With the advent of smartphones, wireless interfaces improved
and computing power increased, thus making it worth looking into
more recent publications.
A lot of work related to device-to-device communication focuses on challenges like
offloading, content dissemination, or energy efficiency
rather than the application layer \cite{AhmedSurveySociallyAware2018}.
Yet, there are some projects related to applications
that describe actual implementations or suggestions for implementations.
In \cite{LuNetworkingsmartphonesdisaster2016}, the authors
use the WiFi ad-hoc mode on an Android device.
This mode is not available by default, an extension had to be compiled into the Linux kernel.
There is still a lack of support of WiFi ad-hoc since publication of that paper (2016),
which shows that only a very limited set of devices would be able to run such an application.
In \cite{YangBlueNetBLEbasedadhoc2017},
the authors use Bluetooth Low Energy (BLE) in IoT scenarios
for ad-hoc communication.
The framework they develop allows devices to communicate without 
predefined roles like client/server.
Other recent works suggest using WiFi Direct
for exchanging data between smartphones
\cite{MaoPerformanceEvaluationWiFi2017,ShuTalk2MeFrameworkDevicetoDevice2018},
which is not readily supported on iOS devices.

Besides the academic work in the field, it is also worth looking into
what related applications are actually available.
In the following, we thus look into existing applications and technologies for
device-to-device data transfer.
Overall, such technologies seem to only operate with devices from the same
manufacturer or the same mobile OS.
Hand-held gaming devices from Nintendo and Sony are offering
data exchange with nearby players in proximity\footnote{\url{https://www.nintendo.com/3ds/built-in-software/streetpass/how-it-works} and \url{http://us.playstation.com/psvita/apps/psvita-app-near.html}}.
This is a feature specific to each gaming system and does not work across devices
from Nintendo and Sony.
Both Apple and Google offer frameworks that enable software developers
to interact with nearby devices.
Apple's framework is called Multipeer Connectivity\footnote{\url{https://developer.apple.com/documentation/multipeerconnectivity}}
and only works with other Apple devices.
Google introduced its framework Google Nearby\footnote{\url{https://developers.google.com/nearby/}} with
two different APIs: Nearby Connections and Nearby Messages.
Nearby Connections allows for device-to-device data transfer, but only between
Android devices.
Nearby Messages is only available when the devices are connected to the internet
and
allows only small payloads, but is available for both Android and iOS.
There are multiplatform apps offering device-to-device data transfer solutions,
e.g., SHAREit\footnote{\url{https://www.ushareit.com/}}.
Such apps often let one user open a WiFi hotspot that
is then joined by a second user.
Data exchange via Bluetooth between Android and iOS devices
is not readily available and typically requires explicit
user interaction.

Reviewing related work in academia, software development frameworks,
and apps, we summarize
that the issue we are facing stems from
the interaction of the following three factors:
\begin{itemize}
	\item \textit{(a) broadcasting}:
	In an optimal solution, exchanging preference data between passing users
	in proximity happens in the background without user interaction.
	\item \textit{(b) multiplatform app}:
	In order to be able to provide the proposed system for virtually all
	smartphone users, the system has to be available on both Android and iOS.
	\item \textit{(c) large message size}:
	In order to exchange user preferences data needed for a recommender
	system, larger messages have to be exchanged (in the range
	of several KB or MB).
\end{itemize}

Solutions for combinations of two of those aspects are available:
(a) + (b): Using Google Nearby Messages, WiFi SSIDs, or Bluetooth Discovery Protocol messages, broadcasting between Android and iOS devices is possible, but only with small payloads.
(b) + (c): Apps like SHAREit allow the explicit connection establishment
between two devices in order to transfer large amounts of data.
Typically, local WiFi hotspots are used.
It might be possible to implement a combination of (a) + (c)
with OS specific solutions (iOS: Multipeer Connectivity, Android: Google Nearby Connections).

\begin{figure}[h]
	\includegraphics[width=1.0\columnwidth]{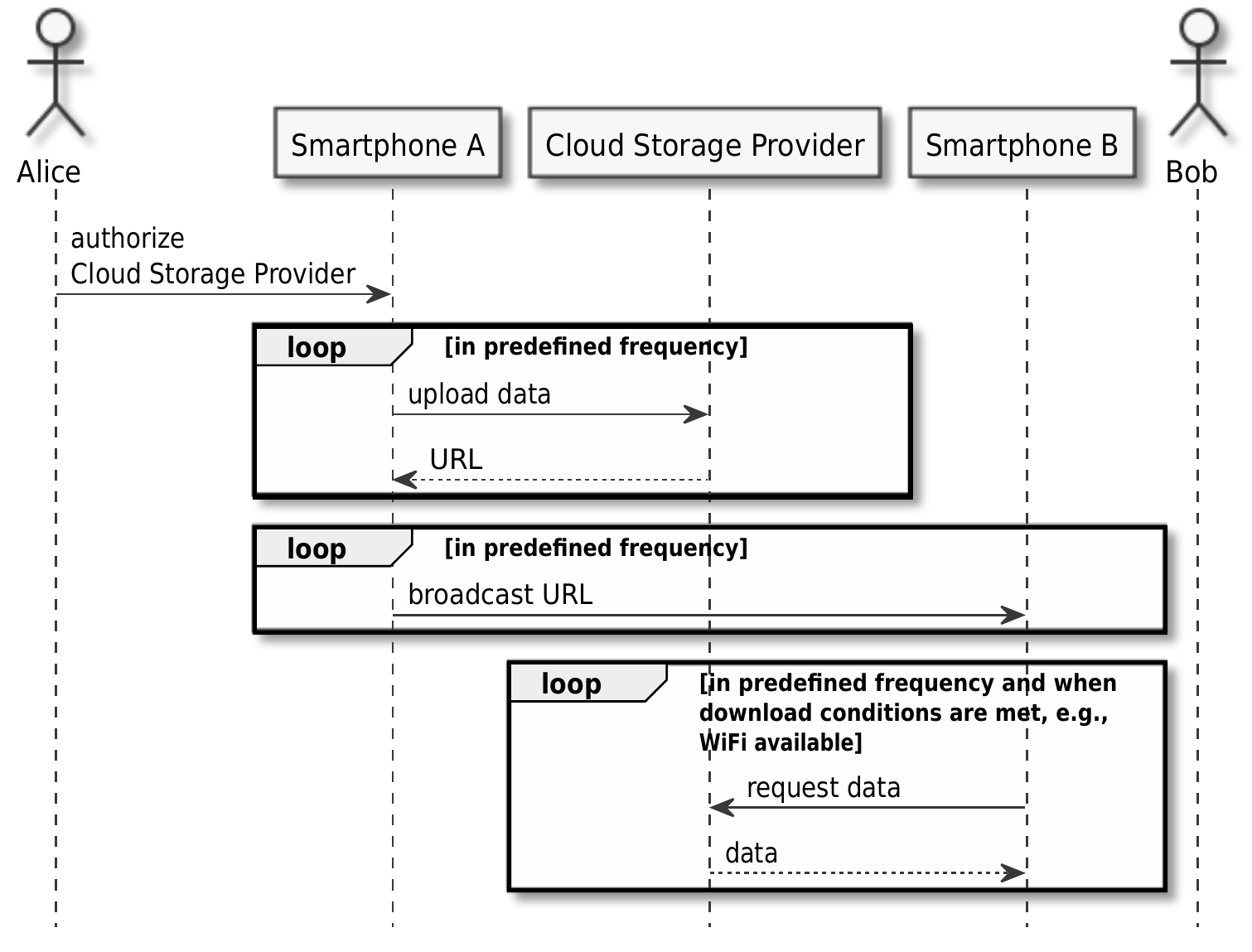}
	\caption{Data exchange via a third-party cloud storage provider.}
	\label{fig:data-exchange}
\end{figure}

For enabling a combination of all three aspects tough, a workaround is necessary.
Building on the existing approaches,
we present one workaround to facilitate a
multiplatform approach with broadcasting that does not require user interaction
and alleviates the issue of size limitations, see Figure
\ref{fig:data-exchange}.

First, Alice authorizes the system to access her account at
some Cloud Storage Provider (CSP) like Dropbox, Google Drive, etc.
Alternatively, she could use her own cloud storage.
In some predefined frequency, Alice's data is then uploaded to the CSP
and shared via a public URL.
This URL is then broadcasted via
Google Nearby Messages or some other technique that allows
multiplatform broadcasting.
As only the URL is shared, which can be further shortened via
a URL shortener service, the small payload restrictions of
multiplatform broadcasting technologies should suffice.
Another user, Bob in Figure \ref{fig:data-exchange},
receives the broadcast with the URL and can
download Alice's publicly shared data.
Optimizations like waiting for a WiFi connection
can easily be implemented.
Note that the only required user interaction by Alice or Bob is the authorization of
the Cloud Storage Provider.
Deeper investigations have to address the limitations posed by such a solution,
for example considering potential attack vectors created by such an approach.

Even though technologies like BLE are in widespread use,
a deeper look reveals that actually exchanging data between
Android and iOS without user interaction is not a trivial task.
Even if future developments might reveal that the implementation
of multiplatform broadcasting gets even harder,
our proposed architecture could still enable ad-hoc group recommendations
with explicit connection establishment.
A group of users in proximity could explicitly enable
the data exchange among the group and locally calculate recommendations
or query a third-party service provider for recommendations.

\section{Recommending new items}
\label{sec:recommending}

Decentralized recommender systems traditionally use peer-to-peer networks
\cite{ZieglerSemanticWebRecommender2005a,Baragliapeertopeerrecommendersystem2013,RuffoPeertoPeerRecommenderSystem2009}.
Gossip protocols leverage commonly encountered small world properties of overlay
topologies in file-sharing networks and allow to find
and gather peers with similar preferences quickly and reliably.
Once established, communities of interest exchange item ratings among each other.
In contrast to such an approach, our ad-hoc fashion of connection and data exchange
between smartphones is a network that is essentially fully disconnected.
Furthermore, most of the given approaches of decentralized recommender systems
deal with personal computers, while we focus on ubiquitous scenarios with mobile
devices.

Other related fields are those of ubiquitous recommender systems
and context-aware recommender systems (CARS).
They consider items in proximity or
consider the user's current context while recommending items, respectively
\cite{MettourisUbiquitousrecommendersystems2014}.
In contrast to these approaches, our proposed system is general in
the sense that any type of item can be recommended,
independent of the item's physical proximity or the user's context.

Another field, that has gained less attention in industry and academia, is that of
group recommender systems \cite{crossen_flytrap:_2002,BeierlePrivacyawareSocialMusic2016}.
With its ad-hoc nature and immediate preference
data exchange, our proposed system
is ideally suited to be used for pervasive group recommendation scenarios.
Exchanging data between several users in a group setting, a local recommender system
can calculate recommendations based on the given data, considering the preferences
of each user.
When utilizing an external service provider for a recommendation,
most likely, before contacting it, the preferences of each group member have to be combined
into one group profile as most providers will only recommend items for a single user.

When employing a local recommender system on the smartphone, additional data is needed.
For content-based filtering,
the properties of items have to be known.
Third-party service providers can help with retrieving such needed metadata about items.
For user-based collaborative filtering, information about the similarity of users is utilized.
While services like Spotify or Netflix have very large databases with millions of users,
the local databases in our proposed architecture will be
much smaller and thus there is a lower likelihood of
finding 
similar users.

We see two possible solutions for this problem.
First, we could let each user disseminate more than just his/her own
item preferences/ratings
and let him/her also send data from previous encounters --
this would also address the cold start problem new users will face.
Another approach is to calculate the similarity of users in a different way,
independent of the users' ratings.
In psychology, the \emph{propinquity effect} is the well-studied effect
that physical proximity is a good predictor of forming interpersonal bonds
\cite{MarvinOccupationalPropinquityFactor1918,Festingerspatialecologygroup1950}.
Having unique identifiers for each user
and counting the number of times
and/or the duration of being in proximity
would then likely predict a higher bond.
Additional methods are available for determining similarity in
proximity-based applications.
In \cite{BeierleYouWhatSimilarity2018},
we developed and evaluated a method for estimating
similarity based on the exchange of the users' context data
using probabilistic data structures
in device-to-device scenarios.
In \cite{EichingeraffinitySystemLatent2019},
we developed a privacy-preserving method for determining the similarity of
two users based on their text messaging data.
Both of those methods can be implemented in our proposed
architecture to find similar users,
without having the need to have users that rated the same items.
Future work will have to show to what extend those similarity metrics
and the propinquity effect
yield valuable similarity indications
for user-based collaborative filtering.
Future work could also investigate the feasibility of approaches
like federated learning, effectively exchanging
trained models or updates to models for recommendations
\cite{McMahanCommunicationEfficientLearningDeep2017}.

\section{Related Work}
\label{sec:relatedwork}

Most related work is from the field of social networking services.
Often, the related work does not specifically focus on recommender
systems but on other aspects like privacy, utilizing
opportunistic networks, or encouraging real-life user interactions.

In our own previous work,
we outlined a general concept for a social networking service
utilizing context data from smartphones, focusing on
how connections between users can be established
\cite{BeierleThreetieredSocialGraph2015}.
In \cite{BeierlePsychometricsbasedFriendRecommendations2017},
we proposed to utilize the relationship between smartphone data and personality traits
for friend recommendations in device-to-device scenarios.

In \cite{BeachWhozThatEvolvingEcosystem2008},
the authors let users sent their identifiers of
a social networking services (via Bluetooth) to enable receivers to visit their publicly
available profile.
The idea is to encourage social interaction.
The authors of \cite{BoutetC3PONetworkApplication2015a}
developed a framework for ephemeral social networks.
The goal is content dissemination in opportunistic networks
in scenarios with large crowds like sports events.
Westerkamp et al.\ propose a decentralized social networking service
in \cite{WesterkampTawkiSelfSovereignSocial2019}.
The goal is avoiding censorship by utilizing the Ethereum Name Service
for identity management and by using self-hosted data storages
-- or trusted third-party service providers -- for each user.
With such a solution, the challenges of lock-in effects and lack of privacy
are tackled and the user is put in control of his/her data.
In our paper, we followed a similar approach, focusing
on a different application domain (recommender systems)
and on pervasive scenarios.

\cite{ZHANGcollaborativemultidevicecomputing2018}
is a short survey paper that highlights the challenges
of device-to-device computing.
Regarding the wireless network interfaces, the authors
consider cellular networks, WiFi, Bluetooth, and NFC.
In our paper, we gave details about the challenges when
implementing data exchange between devices.

Regarding decentralized recommender systems, 
\cite{ZieglerSemanticWebRecommender2005a} and
\cite{BarbosaDistributedUserBasedCollaborative2018}
follows similar approaches
compared to our proposed system.
In \cite{ZieglerSemanticWebRecommender2005a}, decentrally stored data
from the web is used for a recommender system
running on the user's personal computer.
Since that paper's publication (2005),
the development of mobile devices enable mobile and ubiquitous scenarios
depicted in this paper.
In \cite{BarbosaDistributedUserBasedCollaborative2018},
the authors propose that smartphones exchange data in a device-to-device fashion
and calculate their own recommendation
via collaborative filtering.
The focus of that paper is on the recommender algorithm
that is evaluated with a music data set.
For device-to-device communication, WiFi Direct is proposed.

\section{Conclusion and Future Work}
\label{sec:conclusion}

Current recommender systems often
exhibit a lock-in effect for the user
and
are connected to privacy concerns.
Their recommendations are typically
for single users rather than groups
and might be biased according to the interests of the
providing platform.
We proposed a decentralized mobile architecture for recommender systems
that leverages the preferences/ratings from users that are or have been in proximity.
The introduced system runs on the users' smartphones
and utilizes existing external third-party service providers.

The system consists of three main parts,
\emph{data collection}, \emph{data exchange}, and \emph{recommender system}.
We developed three ways to collect data: tracking context data automatically,
retrieving data from third-party service providers, and
letting the user manually rate items.
We highlighted that while short-range wireless transmission technologies
are implemented on all modern smartphones, exchanging larger amounts
of data without user interaction on a system available for
both Android and iOS remains a challenging task.
We proposed a workaround by broadcasting URLs of
cloud storage providers from which the receivers can then download the sender's data.
Regarding the recommendation process,
when considering user-based collaborative filtering,
we looked into ways for determining the similarity between users,
including methods independent of the users' ratings.

Future work
will be to implement and test the proposed device-to-device
data exchange in order to evaluate its reliability,
transmission times in real user scenarios, and battery consumption.
Regarding the recommender system, future work includes
the implementation of a mobile recommender engine.
A simulation with a real data set can help
evaluate the quality of the recommendations that such a system can provide.
Furthermore, potential attack vectors and the overall security of the system
should be investigated.

\section*{Acknowledgment}

We are grateful for the support provided by
Robert Staake,
Jan Pokorski,
Simone Egger, and
Yong Wu.

\bibliographystyle{IEEEtran}

\end{document}